\documentclass{icrc2009}

\usepackage{graphicx}   
\usepackage{caption}    
\usepackage[font=footnotesize]{subfig} 
\usepackage{fixltx2e}
\usepackage{url}

\newcommand{\shorttitle}[1]%
{\markboth{Proceedings of the 31\MakeLowercase{$^{st}$} ICRC, {\L}\'{o}d\'{z} 2009}{#1} }
\newcommand{\etal}{\MakeLowercase{\textit{et al. }}} 

\begin{document}
\title{First results on the search for dark matter in the Sun\\ with the ANTARES neutrino telescope}

\author{
        \IEEEauthorblockN{G.M.A. Lim\IEEEauthorrefmark{2} on behalf of the ANTARES collaboration}\\
        \IEEEauthorblockA{\IEEEauthorrefmark{2}NIKHEF, Science Park 105, P.O. Box 41882, 1009 DB Amsterdam, The Netherlands}
       }

\shorttitle{G.M.A. Lim - ANTARES dark matter results}
\maketitle

\begin{abstract}

The ANTARES collaboration is currently operating the largest neutrino detector in the Northern Hemisphere. One of the goals of ANTARES is the search for dark matter in the universe. In this paper, the first results on the search for dark matter in the Sun with ANTARES in its 5 line configuration, as well as sensitivity studies on the dark matter search with the full ANTARES detector and the future cubic-kilometer neutrino telescope studied by the KM3NeT consortium are presented.

\end{abstract}

 
\section{Introduction}
\vspace*{1mm}

Observational evidence shows that the majority of the matter content of the universe is of non-baryonic nature, befittingly called dark matter. Various extensions of the Standard Model of particle physics provide a well-motivated explanation for the constituents of dark matter: the existence of hitherto unobserved massive weakly interacting particles. A favorite amongst the new particle candidates is the neutralino, the lightest superpartner predicted by supersymmetry, itself a well-motivated extension of the Standard Model. 

In the supersymmetric dark matter scenario, neutralinos have been copiously produced in the beginning of the universe. The expansion of the Universe caused a relic neutralino density in the universe today analogous to the cosmic microwave background. These relic neutralinos could accumulate in massive celestial bodies in the Universe like the Sun, thereby increasing the local neutralino annihilation probability. In the annihilation process new particles would be created, amongst which neutrinos. This neutrino flux could be detectable as a localised emission with earth-based neutrino telescopes like ANTARES.\\[-3mm]

\section{The ANTARES neutrino telescope}
\vspace*{1mm}
 
\begin{figure}[!t]
  \centering
  \includegraphics[width=0.45\textwidth]{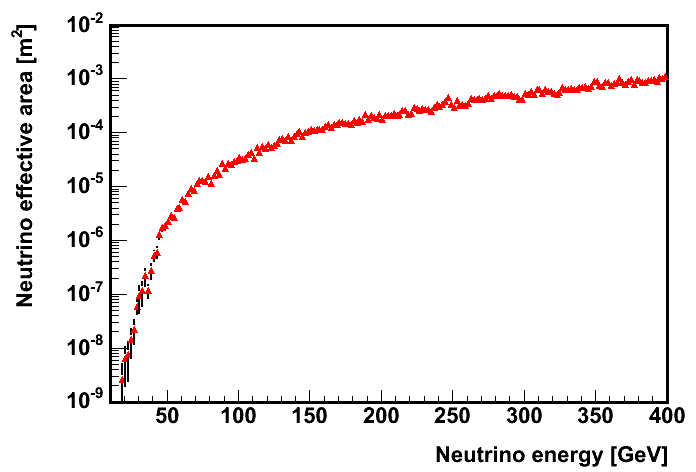}
  \caption{The 12~line ANTARES effective area.}
  \label{NEAnew_colour}
\end{figure} 

\begin{figure*}[!t]
  \centering
  \includegraphics[width=0.8\textwidth]{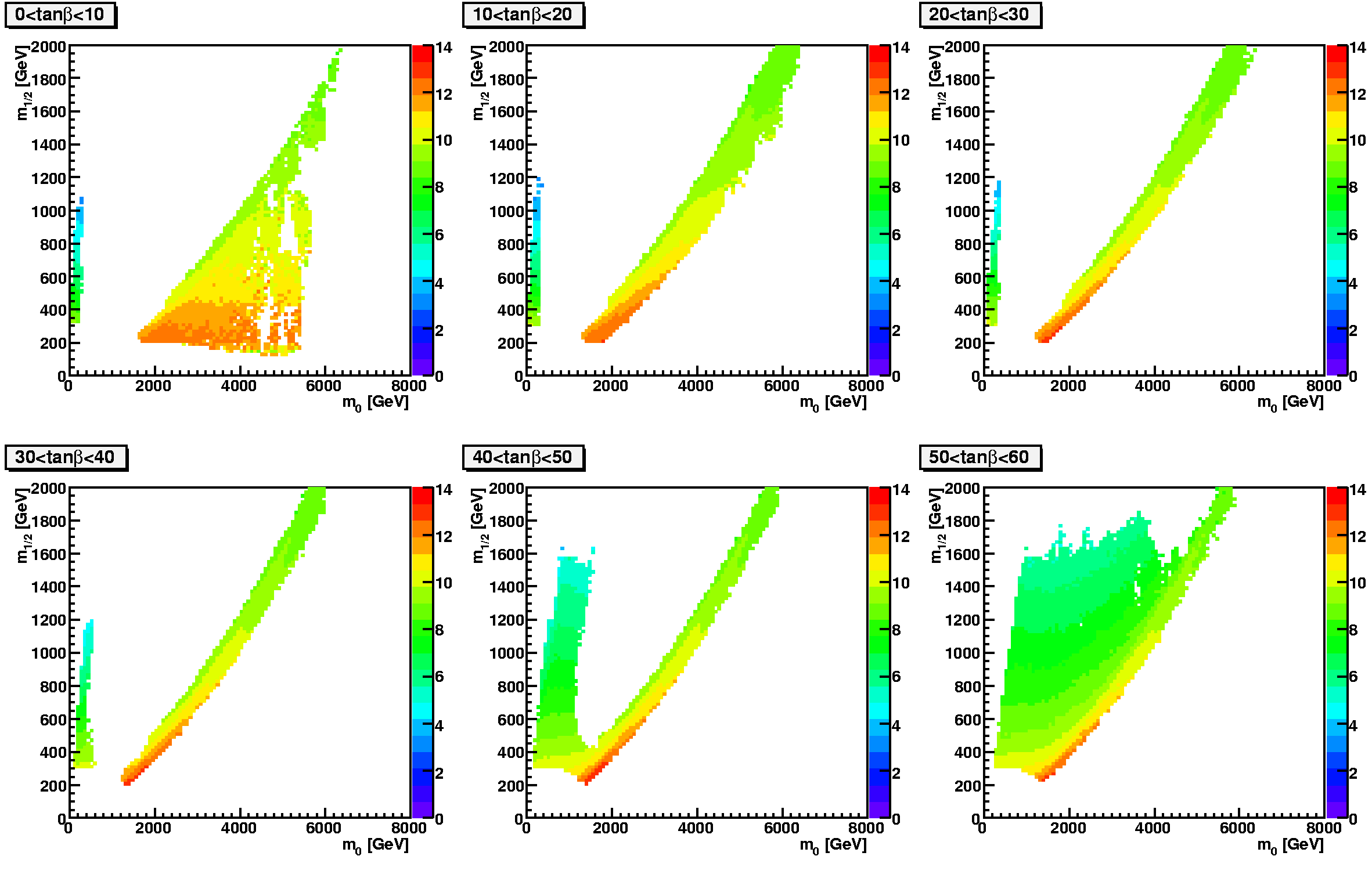}
  \caption{The $\log_{10}(\nu_\mu+\bar{\nu}_\mu)$ flux from neutralino annihilation in the Sun $[\textrm{km}^{-2}\textrm{year}^{-1}]$ in $m_0$-$m_{1/2}$ mSUGRA parameter space, for six different $\tan(\beta)$ intervals. $A_0$ varies between $-3m_0$ and $3m_0$. The flux was integrated above \mbox{$E_\nu=10$~GeV}.}    
  \label{psflux}
\end{figure*}
 
\begin{figure*}[!t]
  \centering
  \includegraphics[width=0.8\textwidth]{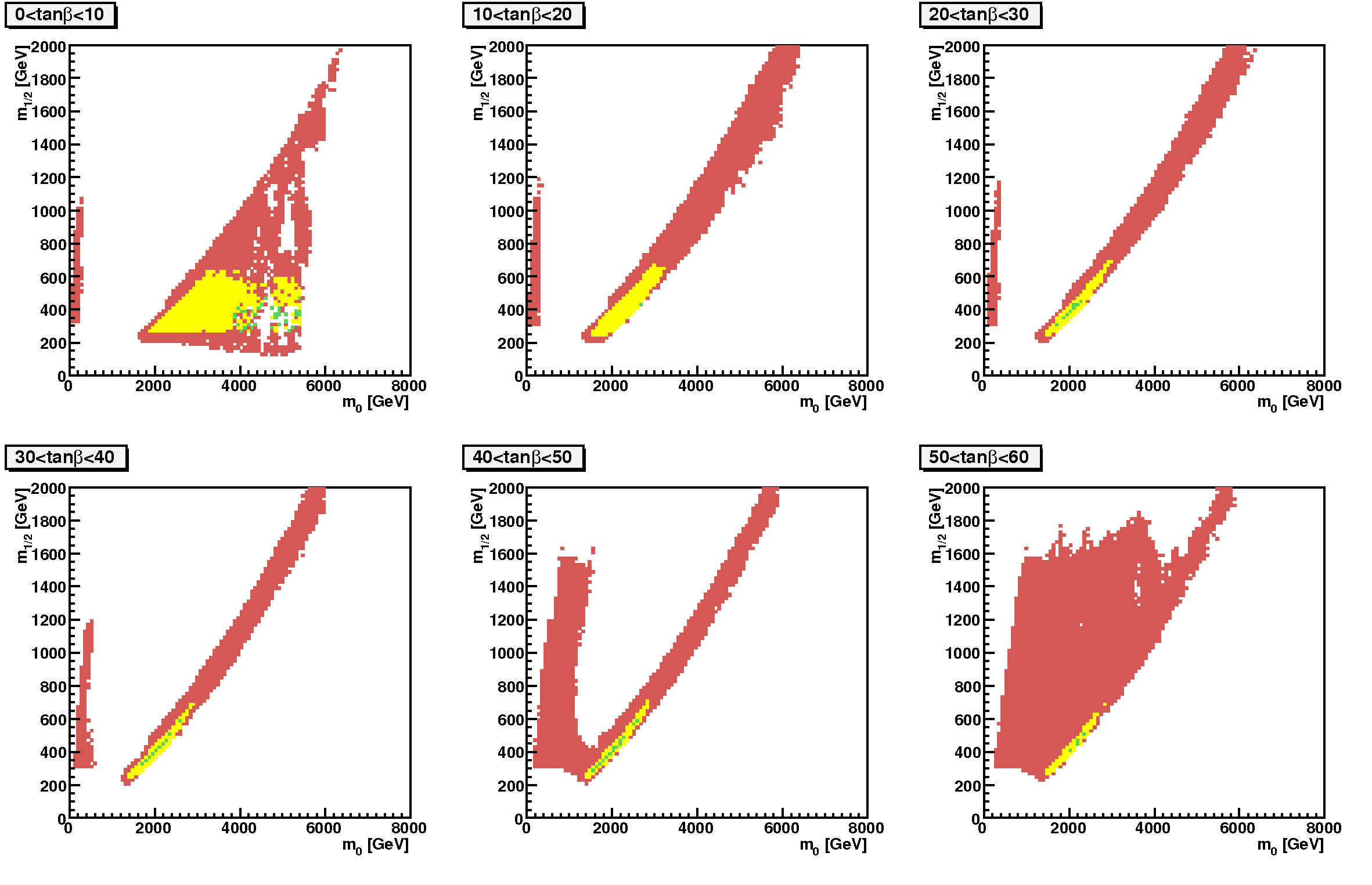}
  \vspace*{-1mm}
  \centerline{\footnotesize (a) ANTARES (12~lines) }\\[4mm]
  \includegraphics[width=0.8\textwidth]{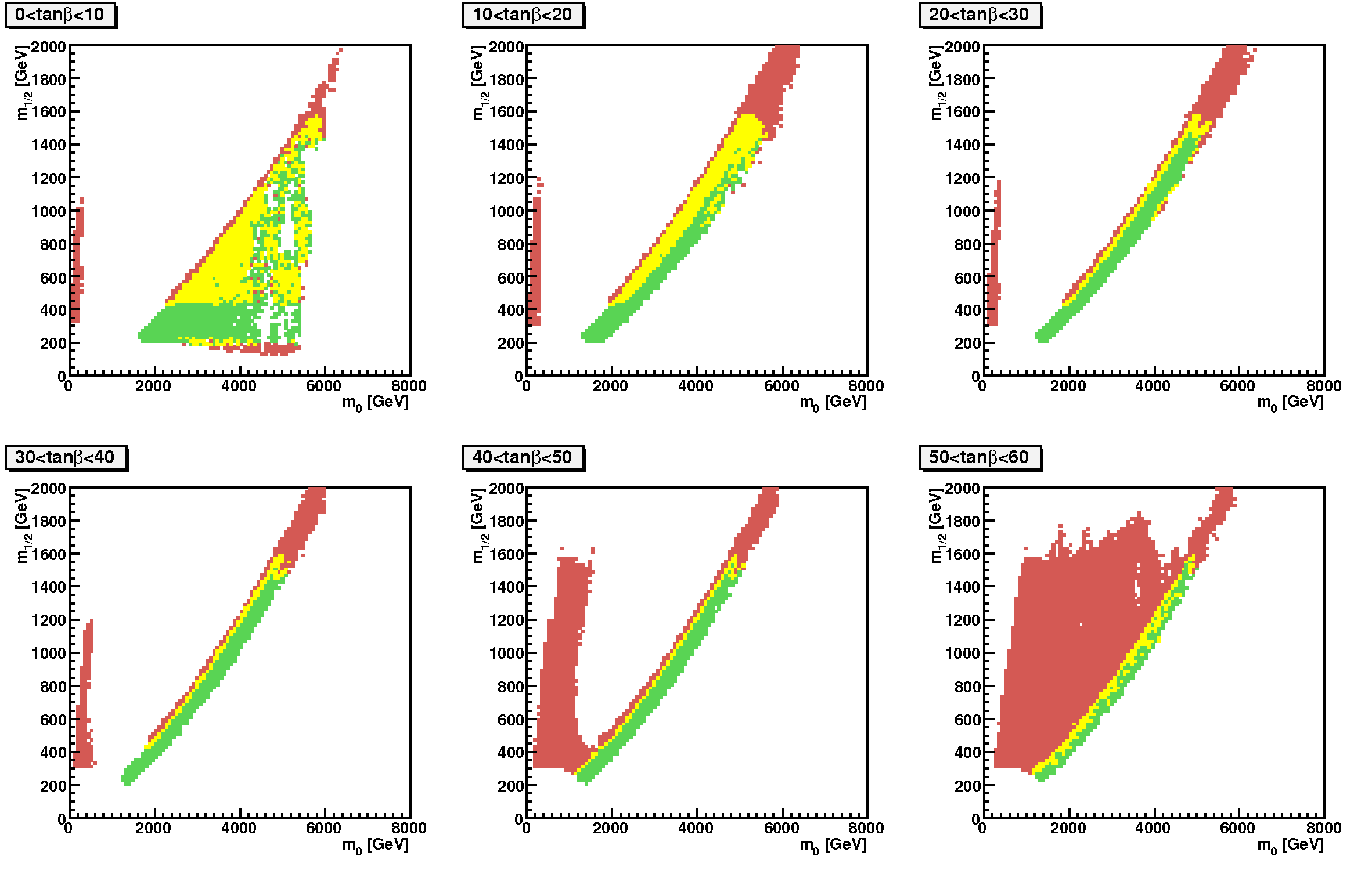}
  \centerline{\footnotesize (b) KM3NeT}\\[2mm]
  \caption{Sensitivity of ANTARES (Fig.~\ref{exclfig}a) and KM3NeT (Fig.~\ref{exclfig}b) in $m_0$-$m_{1/2}$ mSUGRA parameter space, for six different $\tan(\beta)$ intervals. $A_0$ varies between $-3m_0$ and $3m_0$. Green/yellow/red indicates that all/some/no mSUGRA models (depending on~$A_0$ and~$\tan(\beta)$) can be excluded at 90\%~CL after 3~years.}
  \label{exclfig}
\end{figure*}

\addtocounter{figure}{1}

\begin{figure*}[!t]
  \centerline{
    \subfloat[Upper limit on the neutrino flux from the Sun]{\includegraphics[width=0.45\textwidth]{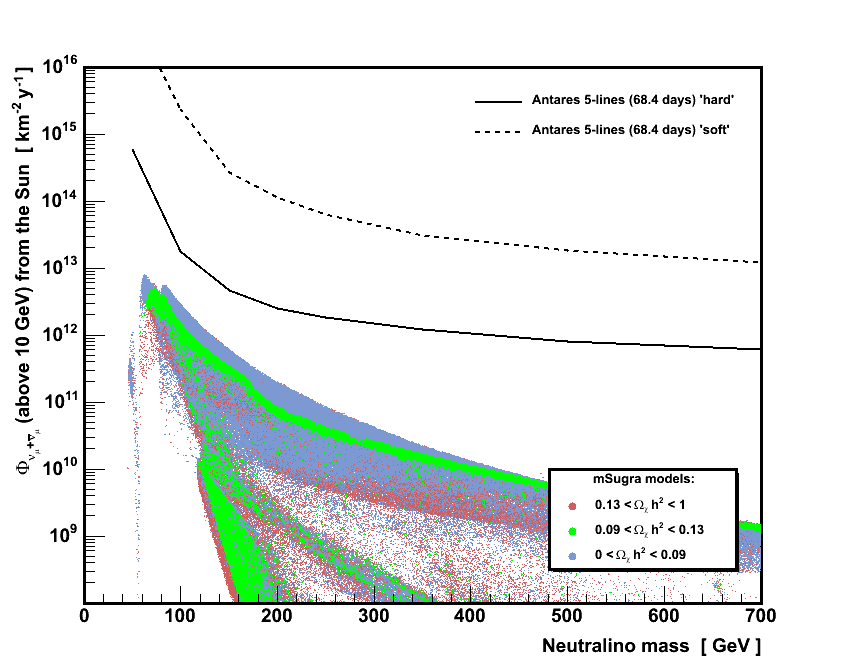} \label{sub_fig1}}
    \hfil
    \subfloat[Upper limit on the muon flux from the Sun]{\includegraphics[width=0.45\textwidth]{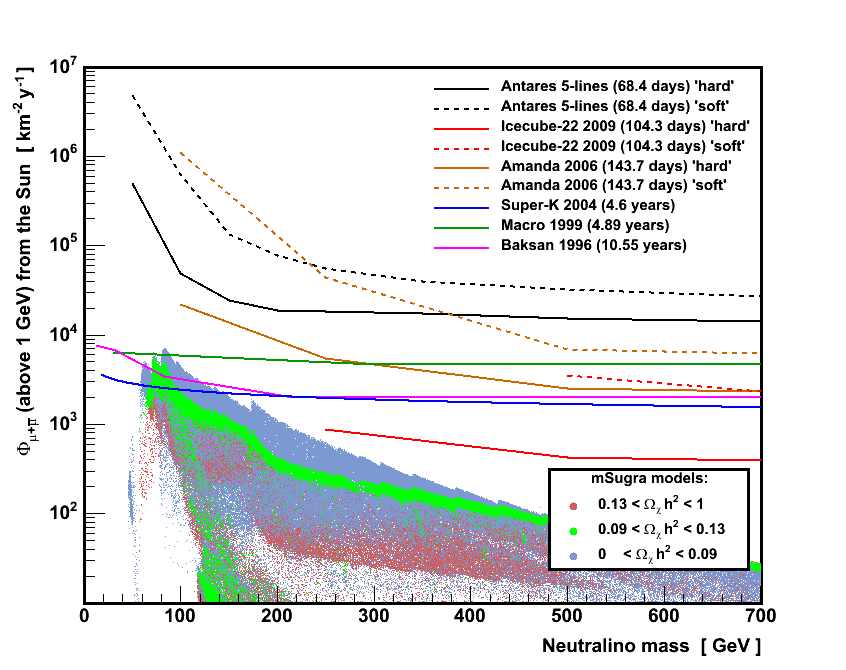} \label{sub_fig2}}
  }
  \vspace*{3mm}
  \caption{Upper limit on the neutrino flux from the Sun above $E_\nu=10$~GeV (Fig.~\ref{sub_fig1}) and the corresponding muon flux above $E_\mu=1$~GeV (Fig.~\ref{sub_fig2}) for the 5-line ANTARES period as a function of the neutralino mass, in comparison to the expected flux from the mSUGRA models considered in Sect. \ref{theory} and other experiments.}
  \label{nflmfl}
\end{figure*}

\addtocounter{figure}{-2}
 
\begin{figure}[!b]
  \centering
  \includegraphics[width=0.45\textwidth]{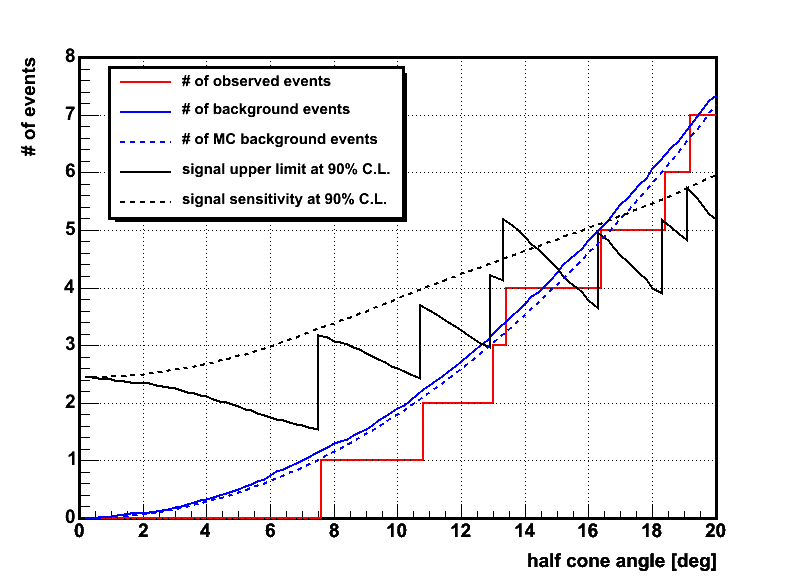}
  \caption{Number of observed neutrinos and expected background events as a function of the search cone radius around the Sun.}
  \label{N_vs_cone_official}
\end{figure}

ANTARES is currently the largest neutrino detector in the Northern Hemisphere~[\ref{antareslabel},\ref{antareslabel2},\ref{antareslabel3}]. Located at a depth of about 2.5~km in the Mediterranean Sea offshore from France, ANTARES is a Cherenkov neutrino telescope comprising 12~detection lines in an approximately cylindrical layout. Each line is comprised of up to 25~storeys, each storey contains \mbox{3~ten-inch~photomultiplier} tubes. The distance between detection lines is 70\,m, and vertically between adjacent storeys 14.5\,m, resulting in an instrumented detector volume of about 0.02~km$^3$. The detector was completed in May~2008. Prior to its completion, ANTARES has been taking data in intermediate configurations of 5~and 10~detector lines for more than one year. The angular resolution of the telescope depends on the neutrino energy~$E_\nu$ and is of the order of one degree at low energy ($E_\nu<$~1~TeV), relevant to dark matter searches.

The sensitivity of a neutrino telescope is conventionally expressed by the neutrino effective area $A_{\textrm{\footnotesize eff}}^{\nu}$. It is defined as

\vspace*{-3mm}
\begin{equation}
  R(t) \;=\; \Phi(E_\nu,t)\;A_{\textrm{\footnotesize eff}}^{\nu}(E_\nu)
  \label{nea}
\end{equation}
\vspace*{-3mm}

\noindent where $R(t)$ is the detection rate and  $\Phi(E_\nu,t)$ is the incoming neutrino flux. The ANTARES effective area (12~lines) for upgoing \mbox{$\nu_\mu+\bar{\nu}_\mu$'s} is shown as a function of the neutrino energy in the low energy regime in Fig.~\ref{NEAnew_colour}. The increase of the effective area with neutrino energy is mainly due to the fact that the neutrino-nucleon cross section as well as the muon range in water are both proportional to the neutrino energy.\\[-3mm]

\section{KM3NeT}
\label{km3net}
\vspace*{1mm}

The KM3NeT consortium aims to build a cubic-kilometer scale neutrino telescope in the Mediterranean Sea~[\ref{KM3NeTlabel},\ref{KM3NeT2label}]. During the Design Study phase of the project, several detector designs are explored. In this study (see Sect.~\ref{sensitivitysect}) we assume one of the possible detector configurations, the so-called ``reference detector''. This homogeneous cubic configuration consists of 225~(15$\times$15)~detection lines, each line carrying 37~optical modules. Each optical module contains \mbox{21~three-inch~photomultiplier} tubes. The distance between detection lines is 95\,m, and vertically between adjacent optical modules 15.5\,m, resulting in an instrumented detector volume of 1~km$^3$.\\[-3mm]

\section{Neutralino annihilation in the Sun} 
\label{theory}
\vspace*{1mm}

We calculated the $\nu_\mu+\bar{\nu}_\mu$ flux resulting from neutralino annihilation in the centre of the Sun. Instead of the general supersymmetry scenario, we used the more constrained approach of minimal supergravity (mSUGRA) in which models are characterized by four parameters and a sign:~$m_{1/2}$, $m_0$, $A_0$, $\tan(\beta)$ and sgn$(\mu)$. The calculation was done using the DarkSUSY simulation package~\cite{DarkSUSY} in combination with the renormalisation group evolution package ISASUGRA~\cite{Isasugra}. We assumed a local neutralino halo density of~0.3~GeV/cm$^3$. To investigate specifically those mSUGRA models that possess a relic neutralino density~$\Omega_\chi$ that is compatible with the cold dark matter density~$\Omega_{\chi,\textrm{\footnotesize WMAP}}$ as measured by WMAP~\cite{WMAP}, the neutrino flux was calculated for approximately four million mSUGRA models using a random walk method in mSUGRA parameter space based on the Metropolis algorithm where~$\Omega_\chi$ acted as a guidance parameter~\cite{MarkovChain}. We considered only \mbox{sgn$(\mu)=+1$} models within the following parameter ranges: \mbox{$0<m_0<8000$~GeV,} \mbox{$0<m_{1/2}<2000$~GeV,} \mbox{$-3m_0<A_0<3m_0$} and \mbox{$0<\tan(\beta)<60$.}

The resulting $\nu_\mu+\bar{\nu}_\mu$ flux from the Sun, integrated above a threshold energy of \mbox{$E_\nu=10$~GeV}, can be seen in the \mbox{$m_0$-$m_{1/2}$~plane} for six different $\tan(\beta)$ ranges in Fig.~\ref{psflux}. Models in the so-called focus point region produce the highest solar neutrino flux. In this region of mSUGRA parameter space the neutralino has a relatively large higgsino component and therefore a large neutralino vector-boson coupling. This enhances the neutralino capture rate in the Sun as well as the neutralino annihilation to vector-bosons, resulting in a relatively high neutrino flux with a relatively hard energy spectrum \cite{Nerzi}.\\[-3mm]

\section{Expected detection sensitivity} 
\label{sensitivitysect}
\vspace*{1mm}

The ANTARES detection sensitivity for neutralino annihilation in the Sun was determined by considering the irreducible background from atmospheric neutrinos and an additional 10\% of that flux due to misreconstructed atmospheric muons in a search cone of 3~degree radius around the Sun. Based on the ANTARES effective area in Fig.~\ref{NEAnew_colour}, the average background prediction after 3~years of effective data taking is \mbox{$\sim\!\!$~7} neutrinos. Assuming that only the average background rate will be measured, a 90\%~CL upper limit on the neutrino flux from the Sun for 3~years of effective data taking was derived, according to~\cite{FeldmanCousins}. This can be compared to the ANTARES detection rate from the expected neutrino flux from neutralino annihilation in the Sun. 

The resulting ANTARES detection sensitivity in the $m_0$-$m_{1/2}$ mSUGRA parameter space for six different $\tan(\beta)$ intervals is shown in Fig.~\ref{exclfig}a. The results of the analogous calculations for the KM3NeT detector outlined in Sect.~\ref{km3net} are shown in Fig.~\ref{exclfig}b. Green/yellow/red colours indicate respectively that all/some/no mSUGRA models (depending on~$A_0$ and~$\tan(\beta)$) can be excluded at 90\%~CL after 3~years of effective data taking by the considered experiment. The sensitivity of ANTARES is sufficient to put constraints on parts of the focus point region of mSUGRA parameter space, while KM3NeT would be sensitive to most of this region.\\[-3mm]

\section{Data analysis} 
\vspace*{1mm}

The data taken during the operation of the first 5~lines (Jan-Dec '07) were used to search for a possible excess in the neutrino flux from the Sun. The effective livetime of this period corresponds to 68.4~days, reduced from the full 164~days due to detector dead-time and the condition that the Sun has to be below the horizon. The number of observed neutrinos in a search cone around the Sun is shown in Fig.~\ref{N_vs_cone_official} as a function of the search cone radius. The expected number of background events from Monte Carlo simulation is in good agreement with the estimation obtained by randomising the direction and arrival time of the observed events. The upper limit and sensitivity on the number of signal events according to~\cite{FeldmanCousins} is also shown.

The corresponding upper limit on the neutrino flux from the Sun was calculated as a function of the neutralino mass assuming two extreme annihilation cases: Pure annihilation into vector-bosons and into $b\bar{b}$~quarks only, referred to as ``hard'' and ``soft'' annihilation respectively. The neutrino energy spectra at Earth for both cases were calculated as a function of the neutralino mass using the simulation package WimpSim~\cite{wimpsim}, assuming the standard oscillation scenario. The search cone used in both cases was optimized as a function of the neutralino mass by Monte Carlo simulation before data analysis. The resulting upper limit at 90\%~CL on the neutrino flux from the Sun, integrated above a threshold energy of \mbox{$E_\nu=10$~GeV}, can be seen for both cases in Fig.~\ref{sub_fig1}. Also shown is the expected neutrino flux from the mSUGRA models considered in Sect.~\ref{theory}, divided into three categories according to the compatibility of their $\Omega_{\chi}$ to $\Omega_{\chi,\textrm{\footnotesize WMAP}}$. Models indicated in green lie within $2\sigma$ of the preferred WMAP value, while models in red/blue have a higher/lower relic density.

The corresponding upper limits on the muon flux above a muon energy threshold of \mbox{$E_\mu=1$~GeV} for both annihilation channels is shown in Fig.~\ref{sub_fig2}, as well as the expected muon flux from the mSUGRA models considered in Sect.~\ref{theory} and upper limits determined by other indirect detection experiments~[\ref{icecubelabel},\ref{amandalabel},\ref{superklabel},\ref{macrolabel},\ref{baksanlabel}].\\[-3mm]

\section{Conclusion} 
\vspace*{1mm}

The first upper limits of the ANTARES detector in its intermediate 5~line configuration on the neutrino and muon flux from neutralino annihilation in the Sun have been obtained. Monte Carlo simulations based on the full ANTARES detector and a possible configuration of the future KM3NeT detector show that these experiments are sensitive to the focus point region of the mSUGRA parameter space.\\[-3mm]


\begin{thebibliography}{99}
\vspace*{1mm}

\bibitem{antares}
\label{antareslabel}
J. Aguilar \etal, Astropart. Phys. {\bf 26}, 314 (2006)
\bibitem{antares2}
\label{antareslabel2}
J. Carr, J. Phys.: Conf. Ser. {\bf 136}, 022047 (2008)
\bibitem{antares3}
\label{antareslabel3}
M. Ageron \etal, Astropart. Phys. {\bf 31}, 277 (2009)
\bibitem{KM3NeT}
\label{KM3NeTlabel}
KM3NeT CDR (2008), \mbox{\url{http://www.km3net.org}}
\bibitem{KM3NeT2}
\label{KM3NeT2label}
U. Katz, Nucl. Instr. Meth. A {\bf 602}, 40 (2009)
\bibitem{DarkSUSY}
P. Gondolo \etal, JCAP {\bf 0407}, 8 (2004)
\bibitem{Isasugra}
H. Baer \etal, \url{arXiv:hep-ph/0312045}
\bibitem{WMAP}
WMAP collaboration, Astrophys. J. {\bf 148}, 175 (2003)
\bibitem{MarkovChain}
E. Baltz, P. Gondolo, JHEP {\bf 10} (2004) 52
\bibitem{Nerzi}
V. Bertin \etal, Eur. Phys. J. C {\bf 26}, 111 (2002)
\bibitem{FeldmanCousins}
G. Feldman, R. Cousins, Phys. Rev. D {\bf 57}, 3873 (1998)
\bibitem{wimpsim}
M. Blennow \etal, JCAP {\bf 0108}, 21 (2008)
\bibitem{icecube}
\label{icecubelabel}
R. Abbasi \etal, Phys. Rev. Lett. {\bf 102}, 201302 (2009) 
\bibitem{amanda}
\label{amandalabel}
M. Ackermann \etal, Astropart. Phys. {\bf 24}, 459 (2006) 
\bibitem{superk}
\label{superklabel}
S. Desai \etal, Phys. Rev. D {\bf 70}, 083523 (2004)
\bibitem{macro}
\label{macrolabel}
M. Ambrosio \etal, Phys. Rev. D {\bf 60}, 082002 (1999)
\bibitem{baksan}
\label{baksanlabel}
M. Boliev \etal, Proc. DARK'96 Heidelberg (1996)
\end{thebibliography}
\end{document}